\documentclass[12pt]{article}
\usepackage{lmodern}
\usepackage{lmodern}
\usepackage[T1]{fontenc}
\usepackage[utf8]{inputenc}
\usepackage[letterpaper]{geometry}
\usepackage{color}
\usepackage{booktabs}
\usepackage{units}
\usepackage{mathtools}
\usepackage{bm}
\usepackage{multirow}
\usepackage{amsmath}
\usepackage{amsthm}
\usepackage{amssymb}
\usepackage{setspace}
\usepackage{microtype}
\usepackage[unicode=true,
 bookmarks=false,
 breaklinks=false,pdfborder={0 0 1},backref=section,colorlinks=true]
 {hyperref}
\hypersetup{
 linkcolor=blue,citecolor=blue,urlcolor=blue,hidelinks}

\makeatletter

\providecommand{\tabularnewline}{\\}

\@ifundefined{date}{}{\date{}}
\usepackage{color}
\usepackage{units}
\usepackage{bm}
\usepackage{amsthm}
\usepackage{amsfonts}
\usepackage{enumitem}
\usepackage{caption}
\usepackage{algorithm}
\usepackage{algpseudocode}

\providecommand{\tabularnewline}{\\}

\providecommand{\Bern}{\operatorname{Bern}}
\providecommand{\IG}{\operatorname{IG}}
\providecommand{\Ga}{\operatorname{Gamma}}
\providecommand{\Exp}{\operatorname{Exp}}

\providecommand{\InvGamma}{\operatorname{InvGamma}}
\providecommand{\GIG}{\operatorname{GIG}}
\providecommand{\Beta}{\operatorname{Beta}}

\makeatother

\begin{document}
\title{\textbf{A Symmetric Random Scan Collapsed Gibbs Sampler for Fully
Bayesian Variable Selection with Spike-and-Slab Priors}}
\author{Mengta Chung}
\maketitle
\begin{center}
Department of Management Sciences, Tamkang University, Taiwan 
\par\end{center}

\begin{center}
mtc@gms.tku.edu.tw
\par\end{center}
\begin{abstract}
We introduce a symmetric random scan Gibbs sampler for scalable Bayesian
variable selection that eliminates storage of the full cross-product
matrix by computing required quantities on-the-fly. Data-informed
proposal weights, constructed from marginal correlations, concentrate
sampling effort on promising candidates while a uniform mixing component
ensures theoretical validity. We provide explicit guidance for selecting
tuning parameters based on the ratio of signal to null correlations,
ensuring adequate posterior exploration. The posterior-mean-size selection
rule provides an adaptive alternative to the median probability model
that automatically calibrates to the effective signal density without
requiring an arbitrary threshold. In simulations with one hundred
thousand predictors, the method achieves sensitivity of 1.000 and
precision above 0.76. Application to a genomic dataset studying riboflavin
production in Bacillus subtilis identifies six genes, all validated
by previous studies using alternative methods. The underlying model
combines a Dirac spike-and-slab prior with Laplace-type shrinkage:
the Dirac spike enforces exact sparsity by setting inactive coefficients
to precisely zero, while the Laplace-type slab provides adaptive regularization
for active coefficients through a local-global scale mixture. 
\end{abstract}

\section{Introduction}

The spike-and-slab prior establishes a Bayesian framework for variable
selection in high-dimensional regression by modeling predictor inclusion
with latent indicators (Mitchell and Beauchamp, 1988; George and McCulloch,
1993). Despite its theoretical advantages, the practical application
of fully Bayesian inference under this prior faces a scalability issue
(Ishwaran and Rao, 2005; Rocková, 2018; Rocková and George, 2014).
As the posterior distribution spans a model space containing $2^{p}$
configurations, Markov chain Monte Carlo (MCMC) methods for exploring
this space suffer from computational and memory requirements that
grow intractably with the number of predictors $p$. Exact MCMC-based
inference is thus widely deemed intractable in ultrahigh-dimensional
settings (Bhattacharya et al., 2016; Blei et al., 2017; Nishimura
and Suchard, 2022; Rocková and George, 2018; Schrider and Kern, 2018;
Scott and Berger, 2010; Shin et al., 2018; Zanella and Roberts, 2019).

Alternative strategies that trade exactness for scalability have been
developed to cope with this practical limitation. One common approach
is to first screen variables based on marginal associations before
applying Bayesian methods to a reduced set, though this ad-hoc filtering
can distort posterior inference (e.g., Fan and Lv, 2008; Zhou et al.,
2022). Variational approximations offer computational speed by optimizing
a simpler distribution but systematically underestimate posterior
uncertainty (e.g., Carbonetto and Stephens, 2012; Wang et al., 2020).
While stochastic search methods (e.g., Benner et al., 2016; Gayevskiy
and Goddard, 2016) achieve tractability by restricting the model space,
they inherently condition the posterior on those restrictions. Consequently,
inference becomes sensitive to algorithmic tuning parameters, such
as the maximum model size and the number of stored models, which are
fixed computational settings rather than random variables informed
by the data via probability theory. Other successful methods are tailored
to specific domains, such as genomics, where they exploit inherent
low-dimensional correlation structures like banded linkage disequilibrium
matrices. However, these specialized solutions are designed around
the specific properties of genomic linkage disequilibrium matrices,
namely, their banded, block-diagonal structure. Consequently, they
do not generalize to other regression problems where the design matrix
or correlation structure does not share these features (e.g., Erbe
et al., 2012; Lloyd-Jones et al., 2019). Each of these valuable strategies
departs from the goal of exact inference in a fully Bayesian spike-and-slab
model.

This study proposes a symmetric random-scan Gibbs sampler (Roberts
and Sahu, 1997; Levine and Casella, 2006), which targets the exact
posterior for spike-and-slab variable selection while directly addressing
the storage and per-iteration cost barriers. Instead of updating all
$p$ coordinates in a systematic scan, we propose updating only a
small, randomly selected subset of $m\ll p$ coordinates. Our method
employs a symmetric selection scheme where both active and inactive
variables are proposed from the same data-informed distribution, allowing
every update to be an exact Gibbs step that requires no Metropolis--Hastings
correction. The random-scan Gibbs sampler guarantees that the stationary
distribution remains the exact posterior, regardless of the selection
probabilities, as long as each variable has a positive chance of being
chosen.

These design choices yield substantial computational savings in both
time and memory; see Section~\ref{sec:complexity} for a concise
comparison with a full-sweep implementation. Beyond computational
efficiency, this framework offers additional flexibility. Data-informed
proposal weights concentrate computation on the most promising variables
without introducing bias, since the exact Gibbs updates are invariant
to the selection probabilities. We operate on the collapsed marginal
likelihood obtained by analytically integrating out the regression
coefficients, which improves Markov chain mixing. A hierarchical Beta-Bernoulli
prior on the global inclusion probability allows the model to adaptively
learn the sparsity level from the data. The primary trade-off is that
more iterations may be required to achieve adequate exploration of
the model space compared to a full-sweep sampler.

The remainder of this paper is organized as follows. Section 2 introduces
the Bayesian linear model with the spike-and-slab prior, derives the
collapsed posterior distribution, and presents the main sampling updates.
Section 3 describes implementation details, including the data-informed
scan mechanism, on-the-fly computation strategy, and computational
complexity. Section 4 reports simulation results. Section 5 presents
an empirical study on the riboflavin dataset. Section 6 concludes
with a discussion.

\section{Methods}

\subsection{Hierarchical Dirac spike with Laplace-type slab priors}

We consider the linear regression model, 
\begin{equation}
\bm{y}\mid\bm{\beta},\sigma^{2}\sim\mathcal{N}\left(\bm{X}\bm{\beta},\sigma^{2}\bm{I}_{n}\right),
\end{equation}
where $\bm{y}\in\mathbb{R}^{n}$ is the response vector, $\bm{X}\in\mathbb{R}^{n\times p}$
is the design matrix with $n$ observations and $p$ predictors, $\bm{\beta}\in\mathbb{R}^{p}$
is the coefficient vector, $\sigma^{2}$ is the error variance, and
$\bm{I}_{n}$ is the $n\times n$ identity matrix.

A Dirac spike-and-(Laplace-type) slab prior induces sparsity and shrinkage
through a spike-and-slab prior on $\bm{\beta}$ with binary inclusion
indicators 
\[
\bm{z}=\left(z_{1},\dots,z_{p}\right)^{\top}\in\left\{ 0,1\right\} ^{p},
\]
where $z_{j}=1$ indicates that predictor $j$ is included in the
model. The index set $\left\{ 1,\dots,p\right\} $ is separated into
the active set $A=\left\{ j:z_{j}=1\right\} $ with corresponding
coefficient $\bm{\beta}_{A}$ and the inactive set $I=\left\{ j:z_{j}=0\right\} $
with corresponding coefficient $\bm{\beta}_{I}$.

Extending Guan and Stephens (2011) with local-global shrinkage (Carvalho
et al., 2010), we adopt the scale mixture representation of Park and
Casella (2008). The hierarchical priors are 
\begin{align}
\bm{\beta}_{A}\mid\bm{\tau}_{A}^{2},\kappa^{2},\bm{z} & \sim\mathcal{N}\left(\bm{0},\bm{D}_{A}\right),\enskip\bm{D}_{A}=\mathrm{diag}\left(\frac{\tau_{j}^{2}}{\kappa^{2}}\right)_{j\in A},\label{eq:prior_beta_A}\\[4pt]
\bm{\beta}_{I}\mid\bm{z} & =\bm{0},\nonumber \\[4pt]
\tau_{j}^{2}\mid\lambda_{1} & \overset{\text{iid}}{\sim}\Exp\left(\frac{\lambda_{1}^{2}}{2}\right),\enskip j=1,\dots,p,\nonumber \\[4pt]
\kappa^{2} & \sim\Ga\left(a_{\kappa},b_{\kappa}\right),\nonumber \\[4pt]
\sigma^{2} & \sim\InvGamma\left(a_{\sigma},b_{\sigma}\right),\nonumber 
\end{align}
where $\bm{\tau}^{2}=\left(\tau_{1}^{2},\dots,\tau_{p}^{2}\right)^{\top}$
are local scale parameters, $\kappa^{2}$ is a global shrinkage parameter,
and $\lambda_{1}>0$ controls the overall degree of sparsity. For
the sparsity, we use 
\begin{align}
z_{j}\mid\pi & \overset{\text{iid}}{\sim}\Bern\left(\pi\right),\enskip j=1,\dots,p,\label{eq:prior_z}\\
\pi\mid a_{\pi},b_{\pi} & \sim\Beta\left(a_{\pi},b_{\pi}\right),\nonumber \\
a_{\pi} & \sim\Ga\left(\alpha_{a},\beta_{a}\right),\nonumber \\
b_{\pi} & \sim\Ga\left(\alpha_{b},\beta_{b}\right).
\end{align}
The hyperparameters $\left(\alpha_{a},\beta_{a},\alpha_{b},\beta_{b}\right)$
favor sparse models while allowing the data to inform the effective
sparsity level. The smaller rate parameter $\beta_{b}$ on $b_{\pi}$
encourages larger values of $b_{\pi}$, which in turn favors smaller
$\pi$ and hence sparser models.

For each coefficient, the conditional prior implied by \eqref{eq:prior_beta_A}-\eqref{eq:prior_z}
can be written as 
\begin{align}
\beta_{j}\mid z_{j},\tau_{j}^{2},\kappa^{2} & \sim\left(1-z_{j}\right)\delta_{0}+z_{j}\mathcal{N}\left(0,\frac{\tau_{j}^{2}}{\kappa^{2}}\right),\label{eq:beta_mixture}\\[4pt]
\tau_{j}^{2} & \sim\Exp\left(\frac{\lambda_{1}^{2}}{2}\right),\nonumber 
\end{align}
so that inactive coefficients are exactly zero while active coefficients
have a Gaussian slab whose variance is decomposed as $\frac{\tau_{j}^{2}}{\kappa^{2}}$.
Integrating out $\tau_{j}^{2}$ yields a Laplace-type global-local
slab density for $\beta_{j}$, combining $\ell_{1}$-like heavy tails
with coefficient-specific adaptivity.

For posterior computation, we use a scale-mixture augmentation. With
the reciprocal local precision $\omega_{j}=\tau_{j}^{-2},$ the conditional
density of $\omega_{j}$ given $\left(\beta_{j},\kappa^{2},\lambda_{1}\right)$
is inverse-Gaussian: 
\[
\omega_{j}\mid\beta_{j},\kappa^{2},\lambda_{1}\sim\IG\left(\mu_{j},\lambda_{1}^{2}\right),
\]
where 
\[
\mu_{j}=\frac{\lambda_{1}}{\left|\beta_{j}\right|\kappa},
\]
and $\IG\left(\mu,\lambda\right)$ denotes the inverse-Gaussian distribution
with mean $\mu$ and shape $\lambda$.

\subsection{Collapsed marginal likelihood}

To avoid repeated $p\times p$ or $n\times n$ factorizations, we
integrate out $\bm{\beta}_{A}$ analytically using the conjugate Gaussian
structure, following the collapsed Gibbs strategy for Bayesian variable
selection (e.g., Bhattacharya, Chakraborty, and Mallick, 2016; Guan
and Stephens, 2011). The marginal likelihood is 
\[
p\left(\boldsymbol{y}|\bm{z},\sigma^{2},\kappa^{2},\boldsymbol{\tau}^{2}\right)=\int p\left(\boldsymbol{y}|\bm{\beta}_{A},\bm{z},\sigma^{2}\right)p\left(\bm{\beta}_{A}|\bm{z},\kappa^{2},\boldsymbol{\tau}^{2}\right)d\bm{\beta}_{A}.
\]
This yields a collapsed likelihood depending only on $\left(\bm{z},\sigma^{2},\kappa^{2},\boldsymbol{\tau}^{2}\right)$,
improving mixing by removing strongly correlated coefficients. We
still sample $\bm{\beta}_{A}$ from its full conditional for posterior
inference.

The collapsed likelihood can be written in terms of the $n\times n$
covariance matrix 
\[
\bm{S}=\sigma^{2}\bm{I}_{n}+\bm{X}_{A}\bm{D}_{A}\bm{X}_{A}^{\top},
\]
so that 
\begin{equation}
\log p\left(\bm{y}|\bm{z},\sigma^{2},\kappa^{2},\bm{\tau}^{2}\right)=-\frac{1}{2}\left(\log|\bm{S}|+\bm{y}^{\top}\bm{S}^{-1}\bm{y}\right)+\text{const}.\label{eq:collapsed_loglik}
\end{equation}
A naive evaluation of \eqref{eq:collapsed_loglik} requires $O\left(n^{3}\right)$
factorizations, but the structure of $\bm{S}$ allows for substantial
savings when $|A|\ll n$.

\subsection{Efficient matrix computation for collapsed likelihood}

To enable efficient computation across different active sets $A$
using the matrix determinant lemma and the Woodbury identity, we precompute
the following auxiliary quantities: 
\begin{align*}
\bm{G}_{0} & =\bm{X}^{\top}\bm{X}\in\mathbb{R}^{p\times p},\\
\bm{h}_{0} & =\bm{X}^{\top}\bm{y}\in\mathbb{R}^{p},\\
c_{y} & =\bm{y}^{\top}\bm{y}\in\mathbb{R}.
\end{align*}
These quantities are computed once at initialization; see Section~\ref{sec:complexity}
for complexity analysis.

During MCMC, when $\sigma^{2}$ changes we use only scalar rescaling.
In particular, we form the scaled length-$p$ vector 
\[
\bm{h}=\frac{1}{\sigma^{2}}\bm{h}_{0}\in\mathbb{R}^{p},
\]
so that $\bm{h}_{A}=\frac{\bm{h}_{0,A}}{\sigma^{2}}$ is immediately
available for any active set $A$. We do not form a full rescaled
$p\times p$ matrix $\frac{\bm{G}_{0}}{\sigma^{2}}$. Instead, whenever
needed we rescale only the active Gram block by 
\[
\bm{G}_{AA}=\frac{1}{\sigma^{2}}\bm{G}_{0,AA}.
\]

With $\bm{D}_{A}$ as defined in \eqref{eq:prior_beta_A}, the active-set
precision matrix is 
\begin{equation}
\bm{M}=\bm{D}_{A}^{-1}+\frac{1}{\sigma^{2}}\bm{X}_{A}^{\top}\bm{X}_{A}=\mathrm{diag}\!\left(\frac{\kappa^{2}}{\tau_{j}^{2}}\right)_{j\in A}+\frac{1}{\sigma^{2}}\bm{G}_{0,AA}\in\mathbb{R}^{|A|\times|A|}.\label{eq:M_def_revised2}
\end{equation}
The matrix $\bm{M}$ is the precision matrix for the active coefficients
and is central to our computational strategy.

Recall $\bm{S}=\sigma^{2}\bm{I}_{n}+\bm{X}_{A}\bm{D}_{A}\bm{X}_{A}^{\top}$.
The Woodbury identity, which has been exploited for fast Bayesian
variable selection (e.g., Bhattacharya et al., 2016; Johndrow, Orenstein,
and Bhattacharya, 2020), states that 
\[
\left(\bm{A}+\bm{U}\bm{C}\bm{V}^{\top}\right)^{-1}=\bm{A}^{-1}-\bm{A}^{-1}\bm{U}\bigl(\bm{C}^{-1}+\bm{V}^{\top}\bm{A}^{-1}\bm{U}\bigr)^{-1}\bm{V}^{\top}\bm{A}^{-1},
\]
applies with $\bm{A}=\sigma^{2}\bm{I}_{n}$, $\bm{U}=\bm{X}_{A}$,
$\bm{C}=\bm{D}_{A}$, and $\bm{V}=\bm{X}_{A}$. Substituting these
choices yields 
\begin{align}
\bm{S}^{-1} & =\sigma^{-2}\bm{I}_{n}-\sigma^{-2}\bm{X}_{A}\left(\bm{D}_{A}^{-1}+\frac{1}{\sigma^{2}}\bm{X}_{A}^{\top}\bm{X}_{A}\right)^{-1}\bm{X}_{A}^{\top}\sigma^{-2}\nonumber \\
 & =\sigma^{-2}\bm{I}_{n}-\sigma^{-2}\bm{X}_{A}\bm{M}^{-1}\bm{X}_{A}^{\top}\sigma^{-2},\label{eq:S_inv_woodbury_revised2}
\end{align}
where $\bm{M}$ is defined in \eqref{eq:M_def_revised2}. Plugging
\eqref{eq:S_inv_woodbury_revised2} into the quadratic form gives
\begin{align}
\bm{y}^{\top}\bm{S}^{-1}\bm{y} & =\sigma^{-2}\bm{y}^{\top}\bm{y}-\sigma^{-4}\bm{y}^{\top}\bm{X}_{A}\bm{M}^{-1}\bm{X}_{A}^{\top}\bm{y}\nonumber \\
 & =\sigma^{-2}c_{y}-\left(\frac{\bm{X}_{A}^{\top}\bm{y}}{\sigma^{2}}\right)^{\top}\bm{M}^{-1}\left(\frac{\bm{X}_{A}^{\top}\bm{y}}{\sigma^{2}}\right)\nonumber \\
 & =\sigma^{-2}c_{y}-\bm{h}_{A}^{\top}\bm{M}^{-1}\bm{h}_{A},\label{eq:quadform_revised2}
\end{align}
where $\bm{h}_{A}=\frac{\bm{h}_{0,A}}{\sigma^{2}}\in\mathbb{R}^{|A|}$.

The \emph{matrix determinant lemma} provides a similar reduction.
For conformable matrices $\bm{A}$, $\bm{U}$, $\bm{C}$, $\bm{V}$
with $\bm{A}$ and $\bm{C}$ invertible, it gives 
\[
|\bm{A}+\bm{U}\bm{C}\bm{V}^{\top}|=|\bm{C}^{-1}+\bm{V}^{\top}\bm{A}^{-1}\bm{U}|\cdot|\bm{C}|\cdot|\bm{A}|,
\]
as applied in scalable spike-and-slab inference (e.g., Carbonetto
and Stephens, 2012; Biswas, Mackey, and Meng, 2022). Applying to $\bm{S}=\sigma^{2}\bm{I}_{n}+\bm{X}_{A}\bm{D}_{A}\bm{X}_{A}^{\top}$
yields 
\begin{align*}
|\bm{S}| & =\left|\bm{D}_{A}^{-1}+\bm{X}_{A}^{\top}\left(\sigma^{2}\bm{I}_{n}\right)^{-1}\bm{X}_{A}\right|\cdot|\bm{D}_{A}|\cdot|\sigma^{2}\bm{I}_{n}|\\
 & =|\bm{M}|\cdot|\bm{D}_{A}|\cdot\left(\sigma^{2}\right)^{n}.
\end{align*}
Taking logarithms and using $|\bm{D}_{A}|=\underset{j\in A}{\prod}\left(\frac{\tau_{j}^{2}}{\kappa^{2}}\right)$,
we obtain 
\begin{equation}
\log|\bm{S}|=n\log\sigma^{2}+\sum_{j\in A}\log\frac{\tau_{j}^{2}}{\kappa^{2}}+\log|\bm{M}|.\label{eq:logdetS_revised2}
\end{equation}
Substituting \eqref{eq:quadform_revised2} and \eqref{eq:logdetS_revised2}
into the collapsed log-likelihood yields a representation depending
only on the $|A|\times|A|$ matrix $\bm{M}$ through $\bm{M}^{-1}$,
$\log|\bm{M}|$, and the quadratic form $\bm{h}_{A}^{\top}\bm{M}^{-1}\bm{h}_{A}$.
This enables efficient updates when $A$ changes by a single inclusion
or exclusion; complexity is analyzed in Section \ref{sec:complexity}.

\subsection{Collapsed Metropolis-within-Gibbs sampler}

\subsubsection{Coordinate-wise inclusion moves}

With the collapsed log-likelihood expressed in terms of $\bm{M}$
and precomputed Gram statistics, we update $\bm{z}$ through coordinate-wise
Metropolis-Hastings moves. At each MCMC iteration, we perform a sequential
scan over coordinates $j=1,\dots,p$. For each $j$, we attempt to
flip $z_{j}$, proposing either an add move $\left(z_{j}:0\to1\right)$
or a drop move $\left(z_{j}:1\to0\right)$. The acceptance probability
depends on the change in the collapsed log-likelihood $\mathcal{L}\left(A\right)$
and the prior odds induced by the Beta-Bernoulli model.

Let $A=\left\{ j:z_{j}=1\right\} $ denote the current active set
and let $A'$ denote the proposed active set obtained by flipping
coordinate $j$. The log-acceptance ratio can be written as 
\[
\log r=\underbrace{\mathcal{L}\left(A'\right)-\mathcal{L}\left(A\right)}_{\text{collapsed likelihood ratio}}+\underbrace{\log p\left(z'_{j}\mid\pi\right)-\log p\left(z_{j}\mid\pi\right)}_{\text{prior ratio}}+\underbrace{\log q\left(A\mid A'\right)-\log q\left(A'\mid A\right)}_{\text{proposal ratio}}.
\]
Because the proposal is a flip and the reverse move is the same flip,
the proposal ratio cancels, 
\[
\log q\left(A\mid A'\right)-\log q\left(A'\mid A\right)=0.
\]
Hence, 
\[
\log r=\mathcal{L}\left(A'\right)-\mathcal{L}\left(A\right)+\log p\left(z'_{j}\mid\pi\right)-\log p\left(z_{j}\mid\pi\right).
\]
Conditional on $\pi$, the Bernoulli$\left(\pi\right)$ prior implies
\[
\log p\left(z'_{j}\mid\pi\right)-\log p\left(z_{j}\mid\pi\right)=\begin{cases}
\log\pi-\log\left(1-\pi\right), & \text{add move }\left(z_{j}:0\to1\right),\\[4pt]
\log\left(1-\pi\right)-\log\pi, & \text{drop move }\left(z_{j}:1\to0\right).
\end{cases}
\]
The MH acceptance probability is 
\[
\alpha=\min\left\{ 1,\exp\left(\log r\right)\right\} ,
\]
implemented by drawing $u\sim\mathrm{Unif}\left(0,1\right)$ and accepting
the proposed flip if $\log u<\log r$.

Each coordinate-wise MH step satisfies detailed balance with respect
to the collapsed target, and the sequential scan preserves the target
distribution while ensuring regular updates for every coordinate.

The main computational task is evaluating $\mathcal{L}\left(A'\right)-\mathcal{L}\left(A\right)$
efficiently. We initialize $\bm{M}^{-1}$ and $\log|\bm{M}|$ once
via Cholesky factorization, then update these quantities under accepted
add/drop moves using Schur-complement-based bordering and downdating
formulas (Ghosh and Clyde, 2011). Proposals leading to a non-positive
Schur complement are skipped for numerical stability.

\subsubsection{Rank-one update formulas}

We derive rank-one update formulas for adding a variable. Consider
adding variable $j\notin A$ to form $A'=A\cup\left\{ j\right\} $.
The prior variance and precision for variable $j$ are $d_{j}=\frac{\tau_{j}^{2}}{\kappa^{2}}$
and $d_{j}^{-1}=\frac{\kappa^{2}}{\tau_{j}^{2}}$. Recall that $\bm{G}=\frac{\bm{G}_{0}}{\sigma^{2}}$
and $\bm{h}=\frac{\bm{h}_{0}}{\sigma^{2}}$, so that 
\begin{align*}
G_{jj} & =\frac{\left(G_{0}\right)_{jj}}{\sigma^{2}},\\
\bm{g} & =\bm{G}_{A,j}=\frac{\bm{G}_{0,Aj}}{\sigma^{2}},\\
h_{j} & =\frac{\left(h_{0}\right)_{j}}{\sigma^{2}}.
\end{align*}
Let $\bm{g}\in\mathbb{R}^{|A|}$ denote the column of $\bm{G}$ corresponding
to interactions between the active variables and variable $j$. The
expanded precision matrix admits the block representation 
\[
\bm{M}'=\begin{bmatrix}\bm{M} & \bm{g}\\
\bm{g}^{\top} & d_{j}^{-1}+G_{jj}
\end{bmatrix}\in\mathbb{R}^{\left(|A|+1\right)\times\left(|A|+1\right)}.
\]
Using the block matrix determinant identity, we obtain 
\begin{equation}
\log|\bm{M}'|=\log|\bm{M}|+\log s,\label{eq:logdet_add_revised}
\end{equation}
where 
\begin{equation}
s=d_{j}^{-1}+G_{jj}-\bm{g}^{\top}\bm{M}^{-1}\bm{g}.\label{eq:schur_s_def}
\end{equation}
The Schur complement $s$ must be positive for numerical validity.

To update the quadratic form, define the residual score 
\begin{equation}
u=h_{j}-\bm{g}^{\top}\bm{M}^{-1}\bm{h}_{A},\label{eq:u_def_revised}
\end{equation}
measuring the association between variable $j$ and the response after
adjusting for the current active set. The inverse of $\bm{M}'$ has
the block form 
\begin{equation}
\left(\bm{M}'\right)^{-1}=\begin{bmatrix}\bm{M}^{-1}+s^{-1}\bm{M}^{-1}\bm{g}\bm{g}^{\top}\bm{M}^{-1} & -s^{-1}\bm{M}^{-1}\bm{g}\\
-s^{-1}\bm{g}^{\top}\bm{M}^{-1} & s^{-1}
\end{bmatrix}.\label{eq:Mprime_inv_revised}
\end{equation}
Let $\bm{h}_{A'}=\left(\bm{h}_{A}^{\top},h_{j}\right)^{\top}\in\mathbb{R}^{|A|+1}$.
Then 
\begin{align}
q\left(A'\right)=\bm{h}_{A'}^{\top}\left(\bm{M}'\right)^{-1}\bm{h}_{A'} & =\bm{h}_{A}^{\top}\bm{M}^{-1}\bm{h}_{A}+s^{-1}\!\left(h_{j}-\bm{g}^{\top}\bm{M}^{-1}\bm{h}_{A}\right)^{2}\nonumber \\
 & =q\left(A\right)+\frac{u^{2}}{s},\label{eq:q_add_revised}
\end{align}
where $q\left(A\right)=\bm{h}_{A}^{\top}\bm{M}^{-1}\bm{h}_{A}$ and
we used \eqref{eq:u_def_revised}.

Combining \eqref{eq:logdet_add_revised} and \eqref{eq:q_add_revised},
and noting that 
\[
\log|\bm{D}_{A'}|-\log|\bm{D}_{A}|=\log d_{j}=\log\frac{\tau_{j}^{2}}{\kappa^{2}},
\]
the change in collapsed log-likelihood under an add move is 
\[
\mathcal{L}\left(A'\right)-\mathcal{L}\left(A\right)=-\frac{1}{2}\left[\log\frac{\tau_{j}^{2}}{\kappa^{2}}+\log s-\frac{u^{2}}{s}\right].
\]
Therefore, the MH log-acceptance ratio for an add move is 
\[
\log r_{\mathrm{add}}=-\frac{1}{2}\left(\log\frac{\tau_{j}^{2}}{\kappa^{2}}+\log s-\frac{u^{2}}{s}\right)+\log\frac{\pi}{1-\pi}.
\]

If the move is accepted, we update $\bm{M}^{-1}$ incrementally using
\eqref{eq:Mprime_inv_revised}. Writing $\bm{t}=\bm{M}^{-1}\bm{g}$,
the accepted-move update is 
\[
\left(\bm{M}'\right)^{-1}=\begin{bmatrix}\bm{M}^{-1}+s^{-1}\bm{t}\bm{t}^{\top} & -s^{-1}\bm{t}\\
-s^{-1}\bm{t}^{\top} & s^{-1}
\end{bmatrix}.
\]
When $A=\varnothing$, $\bm{M}$ is vacuous and $\log|\bm{M}|=0$.
The add-move quantities reduce to 
\begin{align*}
s & =d_{j}^{-1}+G_{jj},\\
u & =h_{j},\\
q\left(A\right) & =0,\\
q\left(A'\right) & =\frac{h_{j}^{2}}{s},
\end{align*}
which is the initialization used before the first variable is included.

For drop moves where $A'=A\setminus\left\{ j\right\} $ with $j\in A$,
analogous rank-one updates can be derived by exploiting the block
structure of $\bm{M}^{-1}$. Without loss of generality, permute the
indices so that $j$ appears last in $A$, and partition 
\[
\bm{M}^{-1}=\begin{bmatrix}\bm{E} & \bm{f}\\
\bm{f}^{\top} & g
\end{bmatrix},
\]
where $g=s^{-1}$, $\bm{E}\in\mathbb{R}^{\left(|A|-1\right)\times\left(|A|-1\right)}$,
$\bm{f}\in\mathbb{R}^{|A|-1}$ and $g\in\mathbb{R}$. The Schur complement
$s$ is the same as in the add move when reintroducing variable $j$
into $A'$, with $g=s^{-1}$. This ensures consistency between add
and drop moves. The determinant update follows from the block determinant
formula. Since $|\bm{M}|=|\bm{M}'|\,s$, we obtain 
\[
\log|\bm{M}'|-\log|\bm{M}|=-\log s.
\]

Next, we update the quadratic form. Partition 
\[
\bm{h}_{A}=\begin{bmatrix}\bm{h}_{-}\\
h_{m}
\end{bmatrix},
\]
where $\bm{h}_{-}\in\mathbb{R}^{|A|-1}$ excludes index $j$, and
$h_{m}$ is the component corresponding to index $j$. The current
quadratic form can be written as 
\[
q\left(A\right)=\bm{h}_{-}^{\top}\bm{E}\bm{h}_{-}+2h_{m}\bm{f}^{\top}\bm{h}_{-}+gh_{m}^{2}.
\]
From the block matrix inversion formula, the inverse of the $\left(|A|-1\right)\times\left(|A|-1\right)$
leading principal submatrix is 
\[
\left(\bm{M}'\right)^{-1}=\bm{E}-g^{-1}\bm{f}\bm{f}^{\top},
\]
and the quadratic form after dropping $j$ becomes 
\[
q\left(A'\right)=\bm{h}_{-}^{\top}\left(\bm{E}-g^{-1}\bm{f}\bm{f}^{\top}\right)\bm{h}_{-}=\bm{h}_{-}^{\top}\bm{E}\bm{h}_{-}-g^{-1}\left(\bm{f}^{\top}\bm{h}_{-}\right)^{2}.
\]
The change in quadratic form is thus 
\[
q\left(A\right)-q\left(A'\right)=2h_{m}\bm{f}^{\top}\bm{h}_{-}+gh_{m}^{2}+g^{-1}\left(\bm{f}^{\top}\bm{h}_{-}\right)^{2}.
\]
From the collapsed log-likelihood $\mathcal{L}\left(A\right)$, only
the terms involving $\frac{\tau_{j}^{2}}{\kappa^{2}}$, $\log|\bm{M}|$,
and $q\left(A\right)$ change when removing variable $j$. Using the
updates above, the change in collapsed log-likelihood is 
\[
\mathcal{L}\left(A'\right)-\mathcal{L}\left(A\right)=-\frac{1}{2}\left[-\log\frac{\tau_{j}^{2}}{\kappa^{2}}-\log s+2h_{m}\bm{f}^{\top}\bm{h}_{-}+gh_{m}^{2}+g^{-1}\left(\bm{f}^{\top}\bm{h}_{-}\right)^{2}\right].
\]
The MH log-acceptance ratio for a drop move is therefore 
\[
\log r_{\text{drop}}=-\frac{1}{2}\left[-\log\frac{\tau_{j}^{2}}{\kappa^{2}}-\log s+2h_{m}\bm{f}^{\top}\bm{h}_{-}+gh_{m}^{2}+g^{-1}\left(\bm{f}^{\top}\bm{h}_{-}\right)^{2}\right]+\log\frac{1-\pi}{\pi}.
\]

\subsubsection{Full sampling cycle}

The full sampling cycle updates the remaining parameters. For the
local shrinkage parameters $\tau_{j}^{2}$, when $j\in A$ is active,
combining the Gaussian prior $\beta_{j}\mid\tau_{j}^{2},\kappa^{2}\sim\mathcal{N}\left(0,\frac{\tau_{j}^{2}}{\kappa^{2}}\right)$
and exponential mixing prior $\tau_{j}^{2}\sim\Exp\left(\frac{\lambda_{1}^{2}}{2}\right)$
gives conditional posterior kernel (writing $x=\tau_{j}^{2}$), 
\begin{align}
p\left(x\mid\beta_{j},\kappa^{2},\lambda_{1}\right)\propto x^{-1/2}\exp\left\{ -\frac{1}{2}\Big(\lambda_{1}^{2}x+\frac{\kappa^{2}\beta_{j}^{2}}{x}\Big)\right\} \label{eq:tau2-kernel}
\end{align}
for $x>0$. This is a generalized inverse Gaussian (GIG) distribution
with 
\[
\tau_{j}^{2}\big|\beta_{j},\kappa^{2},\lambda_{1}\sim\GIG\left(p=\tfrac{1}{2},a=\lambda_{1}^{2},b=\kappa^{2}\beta_{j}^{2}\right).
\]
Working with the reciprocal $\omega_{j}=\frac{1}{\tau_{j}^{2}}$,
\eqref{eq:tau2-kernel} implies 
\[
p\left(\omega_{j}\mid\beta_{j},\kappa^{2},\lambda_{1}\right)\propto\omega_{j}^{-3/2}\exp\left[-\frac{1}{2}\Big(\kappa^{2}\beta_{j}^{2}\omega_{j}+\frac{\lambda_{1}^{2}}{\omega_{j}}\Big)\right]
\]
for $\omega_{j}>0$, which is exactly the inverse-Gaussian (IG) family
with the standard parameterization 
\[
\omega_{j}\big|\beta_{j},\kappa^{2},\lambda_{1}\sim\IG\left(\mu_{j}=\frac{\lambda_{1}}{|\beta_{j}|\kappa},\lambda_{j}=\lambda_{1}^{2}\right).
\]
The local variance is $\tau_{j}^{2}=\omega_{j}^{-1}$. The IG density
$f\left(x|\mu,\lambda\right)\propto x^{-3/2}\exp\left[-\tfrac{\lambda}{2\mu^{2}}x-\tfrac{\lambda}{2}\left(\frac{1}{x}\right)\right]$
matches with $\lambda=\lambda_{1}^{2}$ and $\mu=\frac{\lambda_{1}}{|\beta_{j}|\kappa}$.
For inactive $j\in I$, the local scales are drawn from the prior,
\[
\tau_{j}^{2}\big|z_{j}=0\sim\Exp\left(\frac{\lambda_{1}^{2}}{2}\right).
\]

For $\kappa^{2}$, conjugacy gives a Gamma posterior. With $\kappa^{2}\sim\Ga\left(a_{\kappa},b_{\kappa}\right)$
and $\beta_{j}\mid\tau_{j}^{2},\kappa^{2}\sim\mathcal{N}\left(0,\frac{\tau_{j}^{2}}{\kappa^{2}}\right)$
for active $j\in A$, 
\begin{align*}
p\left(\kappa^{2}\mid\bm{\beta}_{A},\bm{\tau}_{A}^{2}\right) & \propto\left(\kappa^{2}\right)^{a_{\kappa}-1}\exp\left(-b_{\kappa}\kappa^{2}\right)\prod_{j\in A}\left(\kappa^{2}\right)^{\nicefrac{1}{2}}\exp\left(-\frac{\kappa^{2}\beta_{j}^{2}}{2\tau_{j}^{2}}\right)\\
 & \propto\left(\kappa^{2}\right)^{a_{\kappa}+|A|/2-1}\exp\left[-\kappa^{2}\left(b_{\kappa}+\frac{1}{2}\sum_{j\in A}\frac{\beta_{j}^{2}}{\tau_{j}^{2}}\right)\right].
\end{align*}
Thus, 
\[
\kappa^{2}\mid\bm{\beta}_{A},\bm{\tau}_{A}^{2}\sim\Ga\left(a_{\kappa}+\frac{|A|}{2},b_{\kappa}+\frac{1}{2}\sum_{j\in A}\frac{\beta_{j}^{2}}{\tau_{j}^{2}}\right).
\]
When $A=\emptyset$, we draw from the prior $\Ga\left(a_{\kappa},b_{\kappa}\right)$.
Larger $\kappa^{2}$ induces stronger global shrinkage of active coefficients
toward zero.

For $\sigma^{2}$, conjugacy gives an inverse-Gamma posterior. Using
precomputed Gram statistics, the sum of squared errors is 
\[
\text{SSE}=c_{y}-2\bm{\beta}_{A}^{\top}\bm{h}_{0,A}+\bm{\beta}_{A}^{\top}\bm{G}_{0,AA}\bm{\beta}_{A},
\]
avoiding explicit computation of $\bm{X}\bm{\beta}$. With prior $\sigma^{2}\sim\InvGamma\left(a_{\sigma},b_{\sigma}\right)$,
the conditional posterior is 
\[
\sigma^{2}\mid\bm{\beta},\bm{y}\sim\InvGamma\left(a_{\sigma}+\frac{n}{2},b_{\sigma}+\frac{\text{SSE}}{2}\right).
\]

For $\bm{\beta}_{A}$, the conditional posterior is Gaussian with
precision matrix 
\begin{equation}
\bm{\Sigma}_{A}^{-1}=\frac{1}{\sigma^{2}}\bm{G}_{0,AA}+\mathrm{diag}\left(\frac{\kappa^{2}}{\tau_{j}^{2}}\right)_{j\in A}.\label{eq:beta_precision}
\end{equation}
We compute a fresh Cholesky factorization $\bm{\Sigma}_{A}^{-1}=\bm{R}^{\top}\bm{R}$.
The posterior mean is 
\[
\bm{m}_{A}=\bm{\Sigma}_{A}\left(\frac{1}{\sigma^{2}}\bm{h}_{0,A}\right),
\]
where $\bm{h}_{0,A}$ is the subvector of $\bm{h}_{0}$ corresponding
to active indices. The sample $\bm{\beta}_{A}$ is then drawn by solving
triangular systems with $\bm{R}$, 
\[
\bm{\beta}_{A}=\bm{m}_{A}+\bm{R}^{-1}\bm{\epsilon},
\]
where $\bm{\epsilon}\sim\mathcal{N}\left(\bm{0},\bm{I}_{|A|}\right)$.

For $\pi$, the hierarchical Beta-Bernoulli structure uses hyperpriors
$a_{\pi}\sim\Ga\left(\alpha_{a},\beta_{a}\right)$ and $b_{\pi}\sim\Ga\left(\alpha_{b},\beta_{b}\right)$.
The prior mean $\mathbb{E}\left(\pi\mid a_{\pi},b_{\pi}\right)=\frac{a_{\pi}}{a_{\pi}+b_{\pi}}$
encodes the expected fraction of active predictors. Let $K=\sum_{j}z_{j}$.
To center expected model size at $k$, set $b_{\pi}=a_{\pi}\left(\frac{p}{k}-1\right)$.

Given the current inclusion vector $\bm{z}$ and hyperparameters $\left(a_{\pi},b_{\pi}\right)$,
the conditional posterior of $\pi$ is conjugate, 
\begin{equation}
\pi\mid\bm{z},a_{\pi},b_{\pi}\sim\Beta\left(a_{\pi}+|A|,b_{\pi}+\left(p-|A|\right)\right).\label{eq:pi_posterior}
\end{equation}
This induces automatic multiplicity correction (Scott and Berger,
2010): as $p$ increases, the prior shrinks $\pi$ unless there is
substantial evidence for many active coefficients. The joint conditional
posterior of $\left(a_{\pi},b_{\pi}\right)$ is explored via MH. The
unnormalized posterior density is 
\begin{equation}
p\left(a_{\pi},b_{\pi}\mid\pi\right)\propto p\left(\pi\mid a_{\pi},b_{\pi}\right)p\left(a_{\pi}\right)p\left(b_{\pi}\right),
\end{equation}
where 
\begin{align}
p\left(\pi\mid a_{\pi},b_{\pi}\right) & =\frac{\Gamma\left(a_{\pi}+b_{\pi}\right)}{\Gamma\left(a_{\pi}\right)\Gamma\left(b_{\pi}\right)}\pi^{a_{\pi}-1}\left(1-\pi\right)^{b_{\pi}-1},\\[2pt]
p\left(a_{\pi}\right) & =\Ga\left(a_{\pi}|\alpha_{a},\beta_{a}\right),\\
p\left(b_{\pi}\right) & =\Ga\left(b_{\pi}|\alpha_{b},\beta_{b}\right).
\end{align}

We perform random-walk MH on the log-scale. Propose $\log a_{\pi}^{\ast}=\log a_{\pi}+\epsilon_{a}$
and $\log b_{\pi}^{\ast}=\log b_{\pi}+\epsilon_{b}$ with $\epsilon_{a},\epsilon_{b}\overset{\text{iid}}{\sim}\mathcal{N}\left(0,\sigma_{\mathrm{prop}}^{2}\right)$.
The log-acceptance ratio is 
\begin{align*}
\log r=\left[\log p\left(\pi\mid a_{\pi}^{\ast},b_{\pi}^{\ast}\right)+\log p\left(a_{\pi}^{\ast}\right)+\log p\left(b_{\pi}^{\ast}\right)-\log a_{\pi}^{\ast}-\log b_{\pi}^{\ast}\right]\\
-\left[\log p\left(\pi\mid a_{\pi},b_{\pi}\right)+\log p\left(a_{\pi}\right)+\log p\left(b_{\pi}\right)-\log a_{\pi}-\log b_{\pi}\right] & ,
\end{align*}
where $-\log a_{\pi}^{(\ast)}-\log b_{\pi}^{(\ast)}$ terms are Jacobians.
Accept with probability $\min\left\{ 1,\exp\left(\log r\right)\right\} $.

This MH step involves only scalar Beta and Gamma densities, thus increasing
the flexibility of the sparsity prior. Rather than fixing $\left(a_{\pi},b_{\pi}\right)$
(i.e., imposing a fixed prior expected model size), the hierarchical
prior on $\left(a_{\pi},b_{\pi},\pi,\bm{z}\right)$ lets the data
inform both the center and the shape of the prior on $\pi$. In turn,
this yields a posterior distribution on the inclusion indicators $\bm{z}$
that automatically adapts to the effective signal density in the regression,
while retaining the desirable multiplicity-adjustment properties of
the Beta-Bernoulli framework.

\subsubsection{Posterior inference}

The sampler yields posterior draws $\left\{ \left(\bm{z}^{(m)},\bm{\beta}^{(m)}\right):m=1,\dots,M\right\} $
from $M$ iterations, where $z_{j}^{(m)}\in\left\{ 0,1\right\} $
indicates whether predictor $j$ is active in iteration $m$. The
posterior inclusion probability (PIP) is 
\begin{equation}
\widehat{\mathrm{PIP}}_{j}=P\left(z_{j}=1\mid\bm{y}\right)\approx\frac{1}{M}\sum_{m=1}^{M}z_{j}^{(m)},\label{eq:pip_def}
\end{equation}
for $j=1,\dots,p$. The vector $\left(\widehat{\mathrm{PIP}}_{1},\dots,\widehat{\mathrm{PIP}}_{p}\right)$
provides a data-informed ranking of predictors that combines the prior
on $\bm{z}$. Standard use of PIPs in Bayesian variable selection
interprets $\sum_{j}\widehat{\mathrm{PIP}}_{j}$ as the posterior
mean model size and uses PIPs to rank and select predictors (e.g.,
Barbieri and Berger, 2004; Bogdan et al., 2011; Scott and Berger,
2010). A common selection rule is the median probability model (Barbieri
and Berger, 2004), which selects variables with $\widehat{\mathrm{PIP}}_{j}\geqq0.5$.
However, when variables are correlated, many variables might have
PIPs just above 0.5, which leads to inclusion of undesirable ones.
In addition, the 0.5 cutoff does not adapt to the data, which ignores
posterior uncertainty. In settings with weak signals or high predictor
correlation, even true signals may have PIPs below 0.5. The median
model could exclude all of them.

We suggest a posterior-mean-size top-$K$ rule. Let $A=\left\{ j:z_{j}=1\right\} $
denote the set of active predictors in a given model and $|A|=\stackrel[j=1]{p}{\sum}z_{j}$
its size. Since each $z_{j}\in\{0,1\}$ is an indicator variable,
$\mathbb{E}\left[z_{j}\mid\bm{y}\right]=P\left(z_{j}=1\mid\bm{y}\right)$.
By linearity of expectation, 
\[
\mathbb{E}\left[|A|\mid\bm{y}\right]=\mathbb{E}\left[\sum_{j=1}^{p}z_{j}\Bigm|\bm{y}\right]=\sum_{j=1}^{p}\mathbb{E}\left[z_{j}\mid\bm{y}\right]=\sum_{j=1}^{p}P\left(z_{j}=1\mid\bm{y}\right),
\]
so the posterior mean model size can be estimated as 
\begin{equation}
\hat{k}=\sum_{j=1}^{p}\widehat{\mathrm{PIP}}_{j}.\label{eq:khat_def}
\end{equation}
To report a single sparse model, use an integer target size 
\begin{equation}
k_{\star}=\max\left\{ 1,\min\left[p,\mathrm{round}\left(\hat{k}\right)\right]\right\} ,\label{eq:kstar_def}
\end{equation}
which truncates the rounded posterior mean model size to $\left[1,p\right]$.
Sort the PIPs in decreasing order 
\[
\widehat{\mathrm{PIP}}_{(1)}\ge\widehat{\mathrm{PIP}}_{(2)}\ge\cdots\ge\widehat{\mathrm{PIP}}_{(p)},
\]
and take the threshold to be the $k_{\star}$-th largest PIP, 
\begin{equation}
\hat{t}=\widehat{\mathrm{PIP}}_{(k_{\star})}.\label{eq:pip_threshold}
\end{equation}
The selected active set is then 
\begin{equation}
\widehat{A}=\left\{ j:\widehat{\mathrm{PIP}}_{j}\ge\hat{t}\right\} .\label{eq:Ahat_def}
\end{equation}
In the absence of ties at $\hat{t}$, this yields $|\widehat{A}|=k_{\star}$.
If ties occur, $|\widehat{A}|$ may exceed $k_{\star}$ slightly.
Finally, we compute $\widehat{\mathrm{PIP}}_{j}$ via (\ref{eq:pip_def}),
evaluate $\hat{k}$ from (\ref{eq:khat_def}), determine $k_{\star}$
and $\hat{t}$ using (\ref{eq:kstar_def})--(\ref{eq:pip_threshold}),
and report $\widehat{A}$ in (\ref{eq:Ahat_def}) as the selected
set of predictors.

\subsection{Scalable computation via symmetric random scan}

The full-sweep sampler visits all $p$ coordinates in each MCMC iteration.
For ultrahigh-dimensional problems (Fan and Lv, 2008), the per-iteration
cost and storage requirements become prohibitive. To address these
barriers while preserving the exact posterior target, we propose a
symmetric random scan Gibbs sampler with data-informed proposal weights.
The selection weights $w_{j}$ are state-independent, computed once
from marginal correlations and fixed throughout MCMC. This avoids
the $p\times p$ Gram matrix entirely and thus reduces per-iteration
cost.

\subsubsection{On-the-fly Gram computation}

To eliminate storage requirements, we precompute only length-$p$
quantities: 
\begin{align}
s_{j} & =\bm{x}_{j}^{\top}\bm{y}=\left(\bm{h}_{0}\right)_{j},\quad j=1,\dots,p,\label{eq:sj_def}\\
t_{j} & =\|\bm{x}_{j}\|^{2}=\left(\bm{G}_{0}\right)_{jj},\quad j=1,\dots,p,\label{eq:tj_def}\\
c_{y} & =\bm{y}^{\top}\bm{y},\label{eq:cy_def}
\end{align}
where $\bm{x}_{j}\in\mathbb{R}^{n}$ is the $j$-th column of $\bm{X}$.

During MCMC, the cross-product between the active set and candidate
variable $j$ is computed on-the-fly: 
\begin{equation}
\bm{G}_{A,j}=\bm{X}_{A}^{\top}\bm{x}_{j}\in\mathbb{R}^{|A|}.\label{eq:GAj_onthefly}
\end{equation}
The active-set Gram matrix $\bm{G}_{0,AA}\in\mathbb{R}^{|A|\times|A|}$
is maintained incrementally. When adding variable $j$, we compute
$\bm{G}_{A,j}$ via \eqref{eq:GAj_onthefly} and extend $\bm{G}_{0,AA}$;
when removing, we delete the row and column.

\subsubsection{Data-informed proposal weights}

Rather than scanning all $p$ coordinates, we scan a random subset
of $m\ll p$ per iteration. Uniform selection would be inefficient
since true signals are a tiny fraction in ultrahigh-dimensional problems.
To focus effort on promising candidates while maintaining validity,
we use the absolute marginal correlation, 
\begin{equation}
\rho_{j}=\frac{\left|s_{j}\right|}{\sqrt{t_{j}\cdot c_{y}}}=\frac{\left|\bm{x}_{j}^{\top}\bm{y}\right|}{\|\bm{x}_{j}\|\cdot\|\bm{y}\|}=\left|\mathrm{cor}\left(\bm{x}_{j},\bm{y}\right)\right|,\label{eq:rho_j_def}
\end{equation}
as data-informed weights. Variables with high $\rho_{j}$ are more
likely to be active.

Ideally, weights would be proportional to PIPs $\pi_{j}=P\left(z_{j}=1\mid\bm{y}\right)$,
but these are unknown before MCMC. Following the insight of sure independence
screening (Fan and Lv, 2008), we use the marginal correlation $\rho_{j}$
as a proxy for the unknown posterior inclusion probability. Truly
active variables tend to have higher marginal correlations. This motivates
\[
w_{j}^{\mathrm{data}}=\frac{\rho_{j}}{\stackrel[\ell=1]{p}{\sum}\rho_{\ell}}.
\]
However, if $\rho_{j}=0$ for some $j$, then $w_{j}^{\mathrm{data}}=0$,
which violates ergodicity. To ensure $w_{j}>0$ for all $j$, we form
a defensive mixture, 
\begin{equation}
w_{j}=\left(1-\epsilon\right)\cdot\frac{\rho_{j}}{\stackrel[\ell=1]{p}{\sum}\rho_{\ell}}+\frac{\epsilon}{p},\quad j=1,\dots,p,\label{eq:proposal_weights}
\end{equation}
where $\epsilon\in\left(0,1\right)$ (e.g., $\epsilon=0.1$). The
data-informed component focuses updates on promising variables, while
the uniform component guarantees $w_{j}\geq\frac{\epsilon}{p}>0$,
ensuring ergodicity. The normalized weights satisfy $\stackrel[j=1]{p}{\sum}w_{j}=1$.

\subsubsection{Symmetric random scan Gibbs sampler}

Let $m$ be the number of coordinates scanned per iteration, with
$m\ll p$. At each iteration, sample $m$ distinct indices $\left\{ j_{1},\dots,j_{m}\right\} $
from $\left\{ 1,\dots,p\right\} $ without replacement according to
$\left(w_{1},\dots,w_{p}\right)$ in \eqref{eq:proposal_weights}.
Then, for each selected $j$, perform an exact Gibbs update of $z_{j}$
conditional on all other variables.

The idea is that both active and inactive variables are treated symmetrically.
We sample from the same proposal distribution regardless of current
state. This is important for the correct stationary distribution.
An asymmetric random scan would use different selection probabilities
depending on whether a variable is currently active or inactive. For
instance, one might preferentially select active variables for refinement
or focus on inactive variables with high marginal correlations for
potential inclusion. However, such state-dependent selection mechanisms
introduce several complications.

First, asymmetric selection requires proposal ratio corrections in
the acceptance probability. If the probability of selecting variable
$j$ depends on whether $j\in A$, the MH acceptance ratio must include
the factor $\frac{q\left(A\mid A'\right)}{q\left(A'\mid A\right)}$,
where $q\left(\cdot|\cdot\right)$ denotes the state-dependent selection
probability. Computing this correction adds computational overhead
and introduces potential sources of numerical error.

Second, the theoretical analysis becomes substantially more complex.
Verifying detailed balance requires careful bookkeeping of how selection
probabilities change as the active set evolves, and ensuring ergodicity
demands that all possible transitions retain positive probability
under both the forward and reverse selection mechanisms.

Third, asymmetric schemes can introduce subtle biases if the selection
probabilities are not properly balanced. For example, if active variables
are selected more frequently than inactive ones, the chain may exhibit
slow mixing toward sparser models even when the posterior favors them.

The symmetric design eliminates these complications. Because selection
probabilities $w_{j}$ depend only on fixed, precomputed marginal
correlations $\rho_{j}$ and not on the current state $A$, the proposal
ratio $\frac{q\left(A\mid A'\right)}{q\left(A'\mid A\right)}=1$ for
all transitions. Combined with the exact Gibbs update that samples
directly from the full conditional distribution, this yields acceptance
probability 1 for every selected coordinate. The result is a theoretically
clean sampler that preserves the posterior distribution exactly while
concentrating computational effort on promising variables through
the data-informed weights.

For each selected coordinate $j$, we sample $z_{j}$ from its full
conditional distribution: 
\begin{equation}
P(z_{j}=1\mid z_{-j},\bm{y},\bm{\tau}^{2},\kappa^{2},\sigma^{2},\pi)=\frac{1}{1+\exp\left(-\ell_{j}\right)},\label{eq:gibbs_prob}
\end{equation}
where $\ell_{j}$ denotes the log-odds ratio comparing inclusion versus
exclusion of variable $j$, 
\begin{equation}
\ell_{j}=\log\frac{P\left(z_{j}=1\mid\cdot\right)}{P\left(z_{j}=0\mid\cdot\right)}=\mathcal{L}\left(A\cup\{j\}\right)-\mathcal{L}\left(A\setminus\{j\}\right)+\log\frac{\pi}{1-\pi}.\label{eq:log_odds}
\end{equation}
Here, $\mathcal{L}(\cdot)$ denotes the collapsed log-likelihood,
and $\mathcal{L}\left(A\cup\{j\}\right)-\mathcal{L}\left(A\setminus\{j\}\right)$
is computed using the rank-one update formulas developed earlier.

If $z_{j}=0$ (currently inactive), then $j\notin A$, so $A\setminus\{j\}=A$.
The log-odds for adding $j$ is 
\begin{equation}
\ell_{j}=-\frac{1}{2}\left(\log\frac{\tau_{j}^{2}}{\kappa^{2}}+\log s-\frac{u^{2}}{s}\right)+\log\frac{\pi}{1-\pi},\label{eq:log_odds_add}
\end{equation}
where $s$ and $u$ are defined in \eqref{eq:schur_s_def} and \eqref{eq:u_def_revised}.

If $z_{j}=1$ (currently active), then $A\cup\{j\}=A$ and $A\setminus\{j\}$
is the reduced active set, so we evaluate the log-odds for keeping
$j$ versus removing it. Using the drop-move formulae, we obtain 
\begin{equation}
\ell_{j}=-\frac{1}{2}\left(\log\frac{\tau_{j}^{2}}{\kappa^{2}}+\log s-\Delta q\right)+\log\frac{\pi}{1-\pi},\label{eq:log_odds_drop}
\end{equation}
where $\Delta q=2h_{m}\bm{f}^{\top}\bm{h}_{-}+gh_{m}^{2}+g^{-1}(\bm{f}^{\top}\bm{h}_{-})^{2}$
is the change in the quadratic form.

Given $\ell_{j}$, we sample 
\begin{equation}
z_{j}\sim\Bern\left(\frac{1}{1+\exp\left(-\ell_{j}\right)}\right).\label{eq:zj_sample}
\end{equation}
If the sampled value differs from the current state, we update $A$,
$\bm{G}_{0,AA}$, $\bm{h}_{A}$, $\bm{M}^{-1}$, and $\log|\bm{M}|$
accordingly using the rank-one formulae.

\subsubsection{Theoretical justification}

The symmetric random scan Gibbs sampler preserves the posterior distribution
because (i) each single-coordinate Gibbs update satisfies detailed
balance, (ii) convex combinations of invariant kernels remain invariant
(Liu et al., 1995), and (iii) the uniform component $\frac{\epsilon}{p}>0$
ensures ergodicity. Crucially, because the selection weights $w_{j}=\left(1-\epsilon\right)\frac{\rho_{j}}{\underset{\ell}{\sum}\rho_{\ell}}+\frac{\epsilon}{p}$
depend only on precomputed marginal correlations $\rho_{j}=\left|\mathrm{cor}\left(\bm{x}_{j},\bm{y}\right)\right|$,
not on the current state $\bm{z}$, no proposal ratio correction is
needed, and each selected coordinate is updated via exact Gibbs sampling
with acceptance probability 1.

Ergodicity requires that every coordinate has positive selection probability
and that full conditionals have full support. The first condition
holds because $w_{j}\geq\frac{\epsilon}{p}>0$ for all $j$. The second
holds because the Bernoulli full conditional assigns positive probability
to both $z_{j}=0$ and $z_{j}=1$ for any finite log-odds $\ell_{j}$.
Irreducibility follows because any two inclusion vectors $\bm{z},\bm{z}'\in\{0,1\}^{p}$
can be connected by single-coordinate flips, each with positive probability.

\section{Implementation}

\subsection{Computational complexity}

We summarize the computational complexity of both the full-sweep and
symmetric random scan samplers. The auxiliary quantities $\bm{G}_{0}=\bm{X}^{\top}\bm{X}$,
$\bm{h}_{0}=\bm{X}^{\top}\bm{y}$, and $c_{y}=\bm{y}^{\top}\bm{y}$
are computed once at initialization. For dense $\bm{X}$, forming
$\bm{G}_{0}$ scales as $O\left(np^{2}\right)$ and forming $\bm{h}_{0}$
scales as $O(np)$; these one-time costs are amortized over all MCMC
iterations. For the symmetric random scan sampler, the full $\bm{G}_{0}$
matrix is never formed; instead, only the $O\left(p\right)$ diagonal
elements $t_{j}=\|\bm{x}_{j}\|^{2}$ and marginal statistics $s_{j}=\bm{x}_{j}^{\top}\bm{y}$
are precomputed at $O\left(np\right)$ cost with $O\left(p\right)$
storage.

The full-sweep sampler visits all $p$ coordinates per iteration,
yielding per-iteration cost $O\left(p|A|^{2}\right)$ and requiring
$O\left(p^{2}\right)$ storage for $\bm{G}_{0}$. The sum of squared
errors is computed as $\mathrm{SSE}=c_{y}-2\bm{\beta}_{A}^{\top}\bm{h}_{0,A}+\bm{\beta}_{A}^{\top}\bm{G}_{0,AA}\bm{\beta}_{A}$
at $O\left(|A|^{2}\right)$ cost, avoiding explicit computation of
$\bm{X}\bm{\beta}$.

The symmetric random scan sampler has lower per-iteration cost. Sampling
$m$ coordinates from the proposal distribution requires $O\left(m\right)$
or $O\left(p\right)$ operations depending on the sampling algorithm.
For each of the $m$ coordinates, computing $\bm{G}_{A,j}=\bm{X}_{A}^{\top}\bm{x}_{j}$
requires $O\left(n|A|\right)$ operations, evaluating $\ell_{j}$
requires $O\left(|A|^{2}\right)$ , and updating $\bm{M}^{-1}$ if
$z_{j}$ changes requires $O\left(|A|^{2}\right)$ . The Gibbs updates
for the remaining parameters require $O\left(|A|^{3}\right)$ for
the Cholesky factorization in the $\bm{\beta}_{A}$ update and $O\left(|A|\right)$
for $\bm{\tau}^{2}$, $\kappa^{2}$, $\sigma^{2}$, and $\pi$. The
total per-iteration cost is $O\left(mn|A|+m|A|^{2}+|A|^{3}\right)$.
When $|A|\ll n$ and $m\ll p$, this is substantially cheaper than
the full-sweep cost of $O\left(p|A|^{2}\right)$. Since $|A|$ is
typically small in sparse problems (e.g., $|A|\le50$), the cubic
term in $|A|$ remains manageable.

The sampler requires storing the data matrix $\bm{X}\in\mathbb{R}^{n\times p}$
at $O(np)$, the precomputed vectors $\bm{s}=\left(s_{1},\dots,s_{p}\right)^{\top}$
and $\bm{t}=\left(t_{1},\dots,t_{p}\right)^{\top}$ at $O(p)$, the
proposal weights $\left(w_{1},\dots,w_{p}\right)$ at $O(p)$, and
the active-set Gram matrix $\bm{G}_{0,AA}$ together with the precision
inverse $\bm{M}^{-1}$ at $O\left(|A|^{2}\right)$. Crucially, the
$p\times p$ Gram matrix $\bm{G}_{0}$ is never formed or stored.

In sum, the per-iteration cost is reduced from $O\left(p|A|^{2}\right)$
to $O\left(mn|A|+|A|^{3}\right)$, and storage is reduced from $O\left(p^{2}\right)$
to $O\left(np\right)$. For $p=10^{5}$ and $n=500$, the full-sweep
sampler requires approximately 80 GB to store $\boldsymbol{G}_{0}$,
whereas the symmetric random scan sampler requires only 400 MB.

\subsection{Numerical stability}

Our implementation incorporates numerical safeguards for robust performance.
In rank-one updates, we require Schur complement safeguard $s>10^{-12}$.
Proposals violating this threshold, which indicates near-collinearity,
are automatically rejected.

For the local scales, we protect against small $|\beta_{j}|$ by setting
$\beta_{j}^{2}\leftarrow\max\left(\beta_{j}^{2},10^{-12}\right)$
before computing $\mu_{j}=\frac{\lambda_{1}}{|\beta_{j}|\kappa}$.
Following Park and Casella (2008), Makalic and Schmidt (2016), and
Li and Lin (2010), we sample from the inverse-Gaussian distribution
using the Michael-Schucany-Haas algorithm (Michael et al., 1976) and
use $\omega_{j}\leftarrow\mu_{j}$ as a fallback if the sampler returns
a non-finite or non-positive value.

For $\bm{\beta}_{A}$ sampling, we compute a fresh Cholesky factorization
of $\bm{\Sigma}_{A}^{-1}$ at each iteration rather than reusing $\bm{M}^{-1}$
from the $\bm{z}$ updates, avoiding accumulated rounding errors from
repeated rank-one updates.

\section{Simulation}

\subsection{Data generation}

We generate data from the linear regression model $\bm{y}=\mathbf{X}\boldsymbol{\beta}+\boldsymbol{\varepsilon}$,
where $\boldsymbol{\varepsilon}\sim\mathcal{N}\left(\mathbf{0},\sigma^{2}\mathbf{I}_{n}\right)$.
The sample size is $n=500$ and the number of predictors is $p\in\{10^{4},10^{5}\}$,
satisfying the ultrahigh-dimensional criterion of Fan and Lv (2008)
with $\log\left(p\right)=O\left(n^{\xi}\right)$ for $\xi\in\left[0.36,0.39\right]$.
The noise variance is fixed at $\sigma^{2}=1$. The true coefficient
vector $\boldsymbol{\beta}\in\mathbb{R}^{p}$ contains $k=10$ nonzero
entries, 
\[
\beta_{j}=\begin{cases}
1, & j=1,\dots,5,\\
-1, & j=6,\dots,10,\\
0, & j=11,\dots,p.
\end{cases}
\]

This configuration yields signal-to-noise ratio $\mathrm{SNR}\in\left[6.5,8.5\right]$
(Ročková and George, 2018) and proportion of variance explained $\mathrm{PVE}\in\left[0.87,0.89\right]$
(Guan and Stephens, 2011), suggesting that the true signals are identifiable.
Given sufficient signal strength, the challenge is whether the proposed
method can identify true predictors in the setting.

The design matrix $\mathbf{X}$ is generated from a block-correlated
Gaussian distribution (see Chaibub Neto et al., 2014; Foroughi Pour
\& Dalton, 2018; Lu and Petkova, 2014). A $p\times p$ covariance
matrix $\boldsymbol{\Sigma}$ with unit variances and block size 20
is created. For any pair of indices $\left(j,k\right)$ in the same
block with $j\neq k$, we set $\Sigma_{jk}=\rho$ with $\rho\in\left\{ 0.3,0.7\right\} $,
representing moderate and strong within-block dependence; for indices
in different blocks we set $\Sigma_{jk}=0$. Given $\boldsymbol{\Sigma}$,
we compute a regularized Cholesky factor $\mathbf{C}$ and generate
$\mathbf{X}=\mathbf{Z}\mathbf{C}$, where the entries of $\mathbf{Z}\in\mathbb{R}^{n\times p}$
are i.i.d. $\mathcal{N}\left(0,1\right)$. If the final jitter value
is $\mathrm{jitter}$, the rows of $\mathbf{X}$ are $\mathcal{N}\left(\mathbf{0},\boldsymbol{\Sigma}+\mathrm{jitter}\,\mathbf{I}_{p}\right)$,
which reduces to $\mathcal{N}\left(\mathbf{0},\boldsymbol{\Sigma}\right)$
when the Cholesky factorization succeeds with $\mathrm{jitter}=0$.
To obtain stable Cholesky factors of nearly singular covariance matrices,
we use an adaptive jitter strategy. Starting from $\mathrm{jitter}=0$,
we attempt $\mathbf{C}=\operatorname{chol}\left(\boldsymbol{\Sigma}+\mathrm{jitter}\,\mathbf{I}_{p}\right)$.
If this fails, we add a small ridge by setting $\mathrm{jitter}=10^{-8}$;
on further failures, $\mathrm{jitter}$ is multiplied by 10 at each
step, up to a maximum of $10^{-3}$. This exponential backoff yields
a successful Cholesky factorization with minimal perturbation of $\boldsymbol{\Sigma}$.

\subsection{Tuning parameters}

The uniform mixing weight ($\epsilon$) and the number of coordinates
scanned per iteration ($m$) are tuning parameters in the symmetric
random scan sampler. Proper scaling of these parameters with $p$
is essential for reliable inference.

The uniform mixing weight $\epsilon$ controls the balance between
efficiency and coverage. Smaller $\epsilon$ concentrates more probability
on high-signal variables but may under-explore low-signal candidates.
The uniform component $\frac{\epsilon}{p}$ ensures ergodicity regardless
of its magnitude, so smaller $\epsilon$ does not compromise theoretical
validity.

The key consideration for $m$ is that true signal variables must
be visited sufficiently often to achieve accurate posterior inclusion
probability estimates. For a true signal variable $j^{*}$ with marginal
correlation $\rho_{j^{*}}$, the selection probability per iteration
is 
\[
w_{j^{*}}=\left(1-\epsilon\right)\cdot\frac{\rho_{j^{*}}}{\stackrel[\ell=1]{p}{\sum}\rho_{\ell}}+\frac{\epsilon}{p}.
\]
The denominator $\underset{\ell}{\sum}\rho_{\ell}$ is dominated by
the $p-k$ null variables. If null variables have typical marginal
correlations of order $\bar{\rho}_{0}$ (arising from sampling variability),
then $\underset{\ell}{\sum}\rho_{\ell}\approx\bar{\rho}_{0}\cdot p$
for large $p$. Consequently, the data-informed component of $w_{j^{*}}$
scales as 
\[
\frac{\rho_{j^{*}}}{\underset{\ell}{\sum}\rho_{\ell}}\approx\frac{\rho_{j^{*}}}{\bar{\rho}_{0}\cdot p}\propto\frac{1}{p}.
\]
Thus, the selection probability for true signals decreases as $p$
grows, even with data-informed weighting.

The expected number of visits to a true signal over the entire MCMC
run is 
\[
V_{j^{*}}=n_{\mathrm{iter}}\times m\times w_{j^{*}}.
\]
To maintain reliable PIP estimation, we require $V_{j^{*}}$ to exceed
a threshold ($V_{j^{*}}\gtrsim1000$ visits). Since $w_{j^{*}}\propto\frac{1}{p}$,
this requires 
\[
n_{\mathrm{iter}}\times m\propto p.
\]

For signal variables, the population marginal correlation with $\bm{y}$
can be computed from the block-correlated design. Since all 10 signals
lie in the first block and their coefficients sum to zero ($\stackrel[j=1]{10}{\sum}\beta_{j}=0$),
the covariance between a signal predictor $x_{j}$ and the response
is 
\[
\mathrm{cov}\left(x_{j},y\right)=\beta_{j}+\rho\sum_{k\neq j,k\le10}\beta_{k}=\beta_{j}+\rho\left(0-\beta_{j}\right)=\beta_{j}\left(1-\rho\right).
\]
The variance of $y$ is $\mathrm{var}(y)=\bm{\beta}^{\top}\bm{\Sigma}\bm{\beta}+\sigma^{2}=10\left(1-\rho\right)+1=11-10\rho$,
using the fact that cross-terms cancel due to the balanced $\pm1$
coefficients. Thus, the signal marginal correlation is 
\[
\rho_{\mathrm{signal}}=\frac{|1-\rho|}{\sqrt{11-10\rho}}.
\]
For $\rho=0.3$, we have $\rho_{\mathrm{signal}}=\frac{0.7}{\sqrt{8}}=0.247$.
For $\rho=0.7$, we have $\rho_{\mathrm{signal}}=\frac{0.3}{\sqrt{4}}=0.15$.

For null variables, the population correlation is zero (the balanced
signal coefficients cause cancellation even for nulls within the signal
block). As sample correlations are approximately $\mathcal{N}\left(0,\frac{1}{n}\right)$,
so $|r|$ follows a half-normal distribution with mean $\sqrt{\frac{2}{\pi n}}$.
With $n=500$, the expected absolute sample correlation is approximately
\[
\bar{\rho}_{\mathrm{null}}\approx\sqrt{\frac{2}{\pi n}}=\sqrt{\frac{2}{500\pi}}\approx0.036.
\]

The sum of correlations is dominated by the $p-k$ null variables:
\[
\sum_{\ell=1}^{p}\rho_{\ell}\approx k\rho_{\mathrm{signal}}+\left(p-k\right)\bar{\rho}_{\mathrm{null}}\approx p\,\bar{\rho}_{\mathrm{null}}
\]
for $p\gg k$. The selection probability for a true signal becomes
\[
w_{j^{*}}=\left(1-\epsilon\right)\cdot\frac{\rho_{\mathrm{signal}}}{p\bar{\rho}_{\mathrm{null}}}+\frac{\epsilon}{p}=\frac{1}{p}\left[\left(1-\epsilon\right)\frac{\rho_{\mathrm{signal}}}{\bar{\rho}_{\mathrm{null}}}+\epsilon\right].
\]
Let $R=\frac{\rho_{\mathrm{signal}}}{\bar{\rho}_{\mathrm{null}}}$
denote the signal-to-null ratio. For $\rho=0.3$, 
\[
R=\frac{0.247}{0.036}=6.9.
\]
For $\rho=0.7$, 
\[
R=\frac{0.15}{0.036}=4.2.
\]
We use $\epsilon=0.1$ throughout. For $\rho=0.3$ with $R=6.9$,
the coefficient is $\left(1-\epsilon\right)R+\epsilon=6.3.$ For $\rho=0.7$
with $R=4.2$, the coefficient is $\left(1-\epsilon\right)R+\epsilon=3.9.$
To ensure $V_{j^{*}}\geq1000$ visits, we require $n_{\mathrm{iter}}\times m\geq\frac{1000p}{c}$,
where $c\in\left\{ 6.3,3.9\right\} $ depending on $\rho$.

For $p=10^{4}$, we use $m=500$ and $n_{\mathrm{iter}}=10000$, which
gives $n_{\mathrm{iter}}\times m=5\times10^{6}.$ The expected visits
for $\rho=0.3$ are 
\[
V_{j^{*}}=5\times10^{6}\times\frac{6.3}{10000}=3150.
\]
The expected visits for $\rho=0.7$ are 
\[
V_{j^{*}}=5\times10^{6}\times\frac{3.9}{10000}=1950.
\]

For $p=10^{5}$, we use $m=1000$ and $n_{\mathrm{iter}}=30000$,
which gives $n_{\mathrm{iter}}\times m=3\times10^{7}.$ The expected
visits for $\rho=0.3$ are 
\[
V_{j^{*}}=3\times10^{7}\times\frac{6.3}{100000}=1890.
\]
The expected visits for $\rho=0.7$ are 
\[
V_{j^{*}}=3\times10^{7}\times\frac{3.9}{100000}=1170.
\]
All settings ensure $V_{j^{*}}>1000$ for reliable PIP estimation.

\subsection{Initial values}

To favor sparsity without starting from an empty model, we generate
sparse initial values as follows. We set the target model size to
$k_{\mathrm{target}}=20$ and draw an initial active set $A^{(0)}\subset\left\{ 1,\dots,p\right\} $
of size $\min\left(k_{\mathrm{target}},p\right)$ uniformly at random.
We then set $z_{j}^{(0)}=1$ for $j\in A^{(0)}$ and $z_{j}^{(0)}=0$
for $j\notin A^{(0)}$, and initialize the corresponding regression
coefficients as $\beta_{j}^{(0)}\sim\mathcal{N}\left(0,0.1^{2}\right)$
for $j\in A^{(0)}$ with $\beta_{j}^{(0)}=0$ otherwise. The local
scales are initialized via $\tau_{j}^{2\,(0)}\sim\mathrm{Exp}\left(\text{rate}=3\lambda_{1}^{2}\right)$,
imposing strong initial shrinkage. The hyperparameters $\left(\pi^{(0)},a_{\pi}^{(0)},b_{\pi}^{(0)}\right)$
are chosen so that the prior expected model size matches $k_{\mathrm{target}}$.
We set $a_{\pi}^{(0)}=1$, $b_{\pi}^{(0)}=a_{\pi}^{(0)}\left(\frac{p}{k_{\mathrm{target}}}-1\right)$
and $\pi^{(0)}=\frac{k_{\mathrm{target}}}{p}$ so that the expectation
of $\pi$ under the initial Beta prior is $\mathbb{E}\left[\pi^{(0)}\right]=\frac{k_{\mathrm{target}}}{p}$
and hence $\mathbb{E}\left(|A|\right)\approx k_{\mathrm{target}}$.
Finally, the remaining global parameters are initialized at $\kappa^{2\,(0)}=1$
and $\sigma^{2\,(0)}=1$.

\subsection{Results}

Table \ref{tab:simulation_results} summarizes the variable selection
performance across 10 independent replicates. We compare two selection
rules: the $\hat{k}$-rule, which selects the $\hat{k}$ variables
with the highest posterior inclusion probabilities (PIPs), where $\hat{k}$
is estimated from the posterior distribution of model size; and the
median model, which includes all variables with PIP exceeding 0.5.

\begin{table}[htbp]
\centering \caption{Variable selection performance }
\label{tab:simulation_results} %
\begin{tabular}{llccccc}
\hline 
$p$  & $\rho$  & Threshold  & TP  & FP  & Sensitivity  & Precision\tabularnewline
\hline 
\multirow{4}{*}{$10^{4}$} & \multirow{2}{*}{0.3} & $\hat{k}$-rule  & 10.0 (0.0)  & 2.8 (0.6)  & 1.000 (0.000)  & 0.783 (0.039)\tabularnewline
 &  & Median  & 10.0 (0.0)  & 0.2 (0.4)  & 1.000 (0.000)  & 0.982 (0.038)\tabularnewline
\cline{2-7} \cline{3-7} \cline{4-7} \cline{5-7} \cline{6-7} \cline{7-7} 
 & \multirow{2}{*}{0.7} & $\hat{k}$-rule  & 10.0 (0.0)  & 2.5 (1.0)  & 1.000 (0.000)  & 0.804 (0.055)\tabularnewline
 &  & Median  & 10.0 (0.0)  & 0.1 (0.3)  & 1.000 (0.000)  & 0.991 (0.029)\tabularnewline
\hline 
\multirow{4}{*}{$10^{5}$} & \multirow{2}{*}{0.3} & $\hat{k}$-rule  & 10.0 (0.0)  & 3.1 (0.3)  & 1.000 (0.000)  & 0.764 (0.017)\tabularnewline
 &  & Median  & 10.0 (0.0)  & 0.0 (0.0)  & 1.000 (0.000)  & 1.000 (0.000)\tabularnewline
\cline{2-7} \cline{3-7} \cline{4-7} \cline{5-7} \cline{6-7} \cline{7-7} 
 & \multirow{2}{*}{0.7} & $\hat{k}$-rule  & 10.0 (0.0)  & 3.1 (0.6)  & 1.000 (0.000)  & 0.765 (0.033)\tabularnewline
 &  & Median  & 10.0 (0.0)  & 0.1 (0.3)  & 1.000 (0.000)  & 0.991 (0.029)\tabularnewline
\cline{3-7} \cline{4-7} \cline{5-7} \cline{6-7} \cline{7-7} 
\end{tabular}

\textcompwordmark{} \raggedright{}{\footnotesize{}{}{}{}{}Values
are means with standard deviations in parentheses. TP = true positives;
FP = false positives; FN = false negatives. }{\footnotesize\par}
\end{table}

The proposed method achieves adequate variable selection performance
across all simulation settings. Both selection rules attain sensitivity
of 1.000, correctly identifying all 10 true predictors in every replicate.
This demonstrates that the collapsed Gibbs sampler with the Dirac
spike and Laplace-type slab prior adequately recovers the true sparse
signal, even in the challenging ultra-high-dimensional regime ($p=10^{5}$)
with strong within-block correlation ($\rho=0.7$).

The median model achieves high precision ($\geq0.982$) across all
settings. For $p=10^{5}$ with $\rho=0.3$, the median model attains
full recovery in all 10 replicates ($\text{TP}=10$, $\text{FP}=0$,
$\text{Precision}=1$), indicating that the posterior inclusion probabilities
concentrate sharply on the true model. Even under strong correlation
($\rho=0.7$), the median model incurs at most 0.1--0.2 false positives
on average, corresponding to precision above 0.98.

In contrast, the $\hat{k}$-rule tends to overestimate the model size
by 2--3 variables ($\hat{k}\approx12.5$--$13.1$ versus the true
$k=10$), resulting in lower precision ($0.76-0.80$). This reflects
the conservatism of the $\hat{k}$-rule: by selecting the top $\hat{k}$
variables regardless of their PIPs, it includes marginal predictors
that fall below the 0.5 threshold. The median model's explicit PIP
threshold provides a natural guard against such false discoveries.

The results demonstrate robustness. Increasing the dimensionality
tenfold from $p=10^{4}$ to $p=10^{5}$ does not degrade selection
accuracy; if anything, the median model performs slightly better at
higher $p$, consistent with theoretical results on posterior concentration
in sparse high-dimensional regression. Similarly, increasing the within-block
correlation from $\rho=0.3$ to $\rho=0.7$ has minimal impact on
performance, despite the fact that correlated designs typically pose
challenges for variable selection due to multicollinearity among predictors
within the same block.

For $p=10^{4}$, the sampler completes in approximately 3.5 minutes
per replicate, scaling to roughly 60-75 minutes for $p=10^{5}$. The
increase from $p=10^{4}$ to $p=10^{5}$ (approximately 17--22$\times$
for a 10$\times$ increase in $p$) reflects both the larger coordinate
scans ($m=1000$ versus $m=500$) and the longer chains ($n_{\mathrm{iter}}=30000$
versus $n_{\mathrm{iter}}=10000$) required to maintain adequate visits
to true signals. Despite this, the method remains computationally
tractable: a full MCMC run with $p=10^{5}$ predictors completes in
roughly one hour on a single core, making the approach practical for
modern high-dimensional applications. The higher variability at $p=10^{5}$
(standard deviations of 12-23 minutes) likely reflects variation in
the number of active variables visited and the associated matrix operations
during sampling.

\section{Empirical Study}

\subsection{Data and Implementation}

We evaluate our method on the riboflavin production dataset, obtained
from the \texttt{hdi} R package. It is a high-dimensional genomics
benchmark, which contains $n=71$ observations of log-transformed
riboflavin yield in \emph{Bacillus subtilis} with $p=4088$ gene expression
predictors ($\frac{p}{n}\approx57.6$).

The design matrix and response are centered and scaled. We compute
marginal association scores $\rho_{j}=\frac{\left|\bm{X}_{j}^{\top}\bm{y}\right|}{\|\bm{X}_{j}\|\|\bm{y}\|}$.
Comparing the top $k_{0}=20$ scores to the remainder gives a signal-to-null
ratio $R=4.11$. With exploration parameter $\epsilon=0.1$, the effective
concentration is $c=\left(1-\epsilon\right)R+\epsilon=3.80$. To ensure
>1000 expected visits to signals, we require $m\geq\frac{1000\cdot p}{c\cdot n_{\mathrm{iter}}}$.
We set $n_{\mathrm{iter}}=60000$, $m=350$ (yielding 19,500 expected
signal visits), a burn-in of 10000 and thin by 2.

The sampler is initialized randomly. An active set $A^{(0)}$ of size
$k_{\mathrm{target}}=20$ is drawn uniformly, with $\beta_{j}^{(0)}\sim\mathcal{N}\left(0,0.1^{2}\right)$
for $j\in A^{(0)}$. Hyperparameters are set to induce sparsity: $a_{\pi}^{(0)}=1$,
$b_{\pi}^{(0)}=\frac{p}{k_{\mathrm{target}}}-1=203.4$, giving $\mathbb{E}\left[\pi^{(0)}\right]\approx0.0049$.

\subsection{Results and Comparison}

Using the $\hat{k}$-rule (selecting the top $\hat{k}=\sum_{j}\mathrm{PIP}_{j}$
variables), we obtain $\hat{k}=5.85$ and select six genes (Table~\ref{tab:results}).
The error variance estimate is $\hat{\sigma}^{2}=0.195$. The top
gene, \texttt{YOAB\_at} (PIP=0.870), shows a strong negative association
and is consistently identified across major prior studies. Genes \texttt{YXLD\_at}
and \texttt{YXLE\_at} are also frequently selected in the literature.
Our method recovers the complete \emph{yxl} operon triad by including
\texttt{YXLF\_at} and assigns non-negligible probability to \texttt{ARGF\_at}
and \texttt{CARB\_at}, which have appeared in fewer previous analyses.

The selected genes form biologically interpretable clusters. Three
genes, \texttt{YXLD\_at}, \texttt{YXLE\_at}, and \texttt{YXLF\_at},
belong to the \emph{yxl} operon, which encodes membrane-associated
regulatory proteins. The co-selection of this complete triad highlights
the method's ability to capture operon-level signals. Genes \texttt{ARGF\_at}
and \texttt{CARB\_at} participate in arginine and pyrimidine biosynthesis,
suggesting a potential metabolic connection to riboflavin production
through nitrogen metabolism.

\begin{table}[htbp]
\centering \caption{Variable selection results for the riboflavin dataset}
\label{tab:results} %
\begin{tabular}{@{}llll@{}}
\toprule 
Gene  & PIP  & $\hat{\beta}$ (SD)  & Prior studies\tabularnewline
\midrule 
\texttt{YOAB\_at}  & 0.870  & $-0.417$ (0.179)  & a, b, c, d, e, f\tabularnewline
\texttt{ARGF\_at}  & 0.409  & $-0.156$ (0.194)  & c\tabularnewline
\texttt{YXLE\_at}  & 0.378  & $-0.181$ (0.241)  & b, c, e, f\tabularnewline
\texttt{YXLD\_at}  & 0.296  & $-0.140$ (0.222)  & a, b, c, d, e, f\tabularnewline
\texttt{YXLF\_at}  & 0.155  & $-0.064$ (0.154)  & e\tabularnewline
\texttt{CARB\_at}  & 0.130  & $-0.047$ (0.124)  & d\tabularnewline
\midrule 
\multicolumn{4}{l}{\textsuperscript{a}Bühlmann et al.\ (2014), \textsuperscript{b}Javanmard
\& Montanari (2014),}\tabularnewline
\multicolumn{4}{l}{\textsuperscript{c}Chichignoud et al.\ (2016), \textsuperscript{d}Bar
et al.\ (2020),}\tabularnewline
\multicolumn{4}{l}{\textsuperscript{e}Guo et al.\ (2021), \textsuperscript{f}Gong
et al.\ (2021).}\tabularnewline
\end{tabular}
\end{table}

\section{Discussion}

This paper develops a scalable Bayesian variable selection framework
for ultra-high-dimensional regression by imposing a symmetric random
scan Gibbs sampler on a Dirac spike-and-slab prior with Laplace-type
shrinkage. We integrate Dirac spike priors, local-global shrinkage,
and collapsed Gibbs sampling into a unified framework. While each
component has appeared in the literature, their combination yields
a sampler with complementary properties. The Woodbury identity, the
Gram matrix precomputing and matrix determinant lemma are used to
reduce complexity of computation.

The symmetric random scan Gibbs sampler extends scalability to the
ultra-high-dimensional regime by avoiding both full coordinate sweeps
and storage of the $p\times p$ Gram matrix. The key insight is that
data-informed proposal weights, constructed from precomputed marginal
correlations $\rho_{j}=|\mathrm{cor}(\bm{x}_{j},\bm{y})|$, concentrate
sampling effort on promising candidates while the uniform mixing component
ensures ergodicity. Because these weights depend only on fixed quantities
rather than the current state, no proposal ratio correction is required,
and the sampler preserves the exact posterior distribution.

The tuning parameter analysis in Section~4 provides a principled
framework for selecting $m$ and $n_{\mathrm{iter}}$. The formula
$m\geq\frac{1000\cdot p}{c\cdot n_{\mathrm{iter}}}$, where $c=\left(1-\epsilon\right)R+\epsilon$
depends on the signal-to-null ratio $R$, ensures that true signals
receive sufficient visits for reliable PIP estimation. This explicit
characterization of the exploration-exploitation trade-off provides
principled guidance for tuning in high-dimensional MCMC.

We propose the $\hat{k}$-rule as an alternative to the median probability
model for reporting a selected set of predictors. The rule selects
the $\hat{k}=\sum_{j}\widehat{\mathrm{PIP}}_{j}$ variables with highest
PIPs, where $\hat{k}$ is the posterior mean model size. This approach
has several useful properties: it is fully determined by the posterior
distribution without requiring an external threshold, it adapts to
the effective signal density in the data, and it provides a natural
summary when many variables have PIPs clustered near an arbitrary
cutoff.

The simulation results reveal that the $\hat{k}$-rule achieves sensitivity
of 1.000 but includes 2--3 additional variables on average, reflecting
its more inclusive nature. In applications where missing true signals
is costly, the $\hat{k}$-rule may be preferred.

The application to the riboflavin dataset provides external validation
of the method. All six genes selected by the $\hat{k}$-rule have
been identified in at least one previous study using alternative high-dimensional
inference methods (Table~\ref{tab:results}), providing external
support for the selected signals and their biological interpretability.

The proposed Bayesian variable selection method uses the symmetric
random scan sampler to achieve scalability through subsampling rather
than distributional approximations, preserving the theoretical guarantees
of the Gibbs sampler while reducing computational cost. In addition,
the $\hat{k}$-rule for PIPs performs well in both simulation and
empirical studies.

\end{document}